\documentclass[a4paper,12pt,preprint,amsmath,
nofootinbib,superscriptaddress]{revtex4}
\usepackage[hypertex]{hyperref}
\usepackage{graphicx}
\usepackage{epstopdf}
\usepackage{bm}
\usepackage{epsfig}
\usepackage{graphics}
\usepackage{xspace}
\usepackage{amsfonts}

\def\bnslash{\bar n\!\!\!\slash}

\def\OMIT#1{}

\newcommand{\nn}{\nonumber}

\newcommand{\bea}{\begin{eqnarray}}
\newcommand{\eea}{\end{eqnarray}}

\newcommand{\mcdot}{\!\cdot\!}

\newcommand{\Gcusp}{\Gamma_{\rm cusp}}
\newcommand{\half}{\frac{1}{2}}

\newcommand{\muj}{\mu_j}

\newcommand{\cO}{{\cal O}}

\newcommand{\la}{\langle}
\newcommand{\ra}{\rangle}

\begin{document}


\title{Factorization and resummation of t-channel single top
quark production}

\vspace*{1cm}

\author{Jian Wang}
\affiliation{Department of Physics and State Key
Laboratory of Nuclear Physics and Technology, Peking
University, Beijing, 100871, China}

\author{Chong Sheng Li\footnote{Electronic
address: csli@pku.edu.cn}}
\affiliation{Department of Physics and State Key
Laboratory of Nuclear Physics and Technology, Peking
University, Beijing, 100871, China}
\affiliation{Center for High Energy Physics, Peking
University, Beijing, 100871, China}

\author{Hua Xing Zhu}
\affiliation{Department of Physics and State Key
Laboratory of Nuclear Physics and Technology, Peking
University, Beijing, 100871, China}

\author{Jia Jun Zhang}
\affiliation{Department of Physics and State Key
Laboratory of Nuclear Physics and Technology, Peking
University, Beijing, 100871, China}
\affiliation{Nuclear Science Division, MS 70R0319, Lawrence Berkeley National Laboratory,
Berkeley, CA 94720, USA}


\begin{abstract}
 \vspace*{0.3cm}
We investigate the factorization and resummation of t-channel single
top (antitop) quark production in the SM at both the Tevatron and
the LHC in SCET. We find that the resummation effects decrease the
NLO cross sections by about $3\%$ at the Tevatron and about $2\%$ at
the LHC. And the resummation effects significantly reduce the
factorization scale dependence of the total cross section. The
transfer momentum cut dependence and other matching scale
dependencies are also discussed. We also show that when our
numerical results for s- and t-channel single top production at the
Tevatron are combined, it is closer to the experimental result than
the one reported in the previous literatures.
\end{abstract}
\maketitle
\newpage


\section{Introduction}
\label{sec:1}

Top quark is an interesting particle in the standard model (SM)
because of its large mass. It may play a special role in the
electroweak symmetry breaking mechanism. Its properties
 such as mass~\cite{:2009ec}, lifetime~\cite{Amsler20081}, spin,
 couplings to other particles and production mechanism deserve to be studied precisely.

The hadronic production of the single top production provides an
important opportunity to study the charged weak current interactions
of the top quark, e.g., the structure of the the $Wtb$
vertex~\cite{Bernreuther:2008ju}. Besides, it is a background for
many new physics processes. However, due to the difficulties to
discriminate its signatures from the large background, it is only
recently that D0~\cite{Abazov:2009ii} and CDF~\cite{Aaltonen:2009jj}
collaborations have observed the single top production at the Tevatron.

In the three main production modes of the top quark, the t-channel
is specially important because of its largest cross section, which
has been extremely studied, including the next-to-leading order (NLO) corrections
in~\cite{Bordes:1994ki,Stelzer:1995mi,Harris:2002md,Sullivan:2004ie,
Campbell:2004ch,Cao:2004ky,Cao:2005pq}.
 Their results show that the NLO
corrections are about $5\%$ and $9\%$ at the LHC and Tevatron,
respectively. Also, parton shower Monte Carlo simulation was
considered in ~\cite{Frixione:2005vw,
Alioli:2009je}. Threshold resummation for this process was
calculated
in~\cite{Kidonakis:2006bu,Kidonakis:2007ej,Kidonakis:2010tc}, where
only large soft and collinear gluon corrections were resummed and
the resummed cross section was expanded up to $\mathcal
{O}(\alpha_s^3)$.

In this paper, we use the framework of soft-collinear effective theory (SCET)
~\cite{Bauer:2000ew,Bauer:2000yr,Bauer:2001ct,Bauer:2001yt,Becher:2006nr}
 to give a resummed cross section of t-channel single top production, which
  contains all contributions from large logarithms in hard,
 jet and soft functions to all orders. There is a strong motivation
  for performing this calculation because the hard functions of
 this process, compared with soft functions, is not small, but
  refs.~\cite{Kidonakis:2006bu,Kidonakis:2007ej,Kidonakis:2010tc} only resum the soft and jet effects to
   next-to-next-to-leading logarithmic (NNLL) order, using the traditional
  method, and leave the hard effects unresummed. In the SCET approach,
   the different scales in a process
  can be separated because the soft and collinear degrees can be decoupled by the redefinition of the fields.
  At each scale, we only need to deal with the relevant degrees of freedom.
  Their dependencies on the scale are
   controlled by the renormalization group (RG) equations,
    and the hard, jet and soft effects can be resummed conveniently.

This paper is organized as follows. In section~\ref{sec:2}, we give the
factorization and resummation formalism of this process in momentum space.
In section~\ref{sec:3}, each factor in the resummed cross section is calculated.
In section~\ref{sec:4}, we present the numerical results for this process at
the Tevatron and LHC, respectively. We conclude in section~\ref{sec:5}.

\section{Factorization and Resummation Formalism}
\label{sec:2} 
\begin{figure}\centering
  \includegraphics[width=0.4\linewidth]{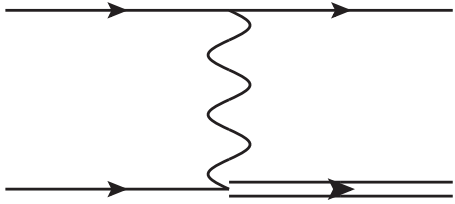}\\
  \caption{The LO Feynman diagram of the t-channel single top production.}
  \label{fig:Born}
\end{figure}

Following the same
factorization formalism as in our previous paper~\cite{Zhu:2010mr},
the partonic differential cross section of t-channel can be written
as
\begin{eqnarray}
\label{main1} \frac{d\hat{\sigma}^{\rm thres}}{d\hat{t}d\hat{u}} &=&
\sum_{ij}\frac{\lambda_{0,ij}}{64\pi N^2_c\hat{s}^2}
\int^{s_4/(2E_1)}_0 dk^+\,
H_{IJ}(\mu)S_{JI}(k^+,\mu)J(s_4-2E_1k^+,\mu),
\end{eqnarray}
with
\begin{equation}
 \lambda_{0,ij} =  \frac{e^4}{\sin^4\theta_W}
|V_{ij}|^2
|V_{tb}|^2 \frac{(\hat{s}-m^2_t)\hat{s}}{(\hat{t}-M^2_W)^2}.
\end{equation}
We use the same definitions of all above notations
as~\cite{Zhu:2010mr} and also choose the independent singlet-octet
basis in color space used in~\cite{Zhu:2010mr}. Because of the
special color structure of this process at leading order (LO), the
hard function matrix elements do not contribute to the cross section
except for $H_{11}$ at the NLO accuracy. In SCET, there is a RG
evolution factor connecting the hard scale $\mu_{h}$ and the final
common scale $\mu$, which would contain contributions from
non-diagonal elements in its anomalous dimension matrix. However,
these contributions are so small, as we saw in
ref.~\cite{Zhu:2010mr}, that we can neglect them. Therefore, we can
consider this channel as a double deep-inelastic-scattering (DDIS)
process~\cite{Harris:2002md} and assume that the hard function
$H_{11}$ can be factorized into two parts, i.e.~$H_{up}$ and
$H_{dn}$, which represent corrections from the up fermion line and down
fermion line in the Feynman diagram~\ref{fig:Born}, respectively. And eq.~(\ref{main1})
can be simplified to
\begin{eqnarray}\label{main2}
\frac{d\hat{\sigma}^{\rm thres}}{d\hat{t}d\hat{u}} =
\sum_{ij}\frac{\lambda_{0,ij}}{64\pi N^2_c\hat{s}^2}
\int^{s_4/(2E_1)}_0 dk^+ H_{up}(\mu) H_{dn}(\mu)
J(s_4-2E_1k^+,\mu)S(k^+,\mu).
\end{eqnarray}

\section{The Hard, Jet and Soft Functions at NLO }
\label{sec:3} At the NNLL accuracy, we need the explicit expressions
of hard, jet and soft functions up to NLO in perturbation theory. In
this section, we show the calculations of them.

\subsection{Hard functions}
\label{subsec:21} The hard functions are the absolute square of the
Wilson coefficients of the operators, which can be obtained by
matching the full theory onto SCET. In practice, we need to
calculate the one-loop on-shell Feynman diagrams of this process in
both the full theory and SCET. In dimensional regularization(DR),
the facts that the IR structure of the full theory and the effective
theory is identical and the on-shell integrals are scaleless and
vanish in SCET imply that the IR divergence of the full theory is
just the negative of the UV divergence of SCET. After calculating
the one-loop on-shell Feynman diagrams, we get the hard functions at
NLO are as follows:
\begin{eqnarray}
\label{hard_function}H_{up}(\mu_{h,up})&=&1+\frac{C_{F}\alpha_{s}(\mu_{h,up})}{4\pi}\biggl(
-2\mathrm{ln}^2\frac{-\hat{t}}{\mu_{h,up}^{2}}+6\mathrm{ln}\frac{-\hat{t}}{\mu_{h,up}^{2}}+c^{H,up}_1 \biggr),\\
\label{hardf_function}H_{dn}(\mu_{h,dn})&=&1+\frac{C_{F}\alpha_{s}(\mu_{h,dn})}{4\pi}\biggl(
-4\mathrm{ln}^2\frac{-t+m_t^2}{\mu_{h,dn}m_t}+10\mathrm{ln}\frac{-t+m_t^2}{\mu_{h,dn}m_t}+c^{H,dn}_1\biggr),
\end{eqnarray}
with
\begin{eqnarray}
c^{H,up}_1&=&-16+\frac{\pi^2}{3},\\
c^{H,dn}_1&=&-\frac{2}{\lambda}\mathrm{ln}(1-\lambda)+2\mathrm{ln}^2(1-\lambda)
+6\mathrm{ln}(1-\lambda)+4\mathrm{Li}_{2}(\lambda)-12-\frac{\pi^2}{6}   \nn\\
&&+\frac{2m_t^2\hat{u}}{\hat{t}(\hat{s}-m_t^2)}\mathrm{ln}\frac{m_t^2}{m_t^2-\hat{t}},
\end{eqnarray}
where $\lambda=\hat{t}/(\hat{t}-m_t^2)$. The hard functions have a
well behaved expansion in powers of the coupling constant, if
 $\mu_{h,up}$ and $\mu_{h,dn}$ are taken to be of order the natural
 scales, $\mu_{h,up} \sim \sqrt{-\hat{t}}$ and $\mu_{h,dn} \sim (-\hat{t}+m_t^2)/m_t$
, respectively. From
eqs.(\ref{hard_function},~\ref{hardf_function}), we can write the RG
equations of hard functions as
\begin{eqnarray}
\frac{d}{d~\mathrm{ln}\mu_{h,up}}H_{up}(\mu_{h,up})&=&\biggl( 2\Gamma_{\rm cusp}
\mathrm{ln}\frac{-\hat{t}}{\mu_{h,up}^2}+2\gamma_{up}^V \biggr )H_{up}(\mu_{h,up}),\\
\frac{d}{d~\mathrm{ln}\mu_{h,dn}}H_{dn}(\mu_{h,dn})&=&
\biggl(2\Gamma_{\rm
cusp}~\mathrm{ln}\frac{-\hat{t}+m_t^2}{\mu_{h,dn}m_t}+2\gamma_{dn}^V
\biggr )H_{dn}(\mu_{h,dn}),
\end{eqnarray}
where $\Gamma_{\rm cusp}$ is related to the cusp anomalous dimension
of Wilson loops with light-like segments~\cite{Korchemskaya1992169},
while $\gamma_{up}^V$ and $\gamma_{dn}^V$ accounts for single-logarithmic evolution.
Their expressions are shown in appendix A.

After solving the RG equations, we get the hard functions at an
arbitrary scale $\mu$:
\begin{eqnarray}
H_{up}(\mu)&=& \mathrm{exp}\bigl[ 4S(\mu_{h,up},\mu)-2a_{up}^V(\mu_{h,up},\mu)\bigr]\bigg( \frac{-\hat{t}}{\mu_{h,up}^2} \bigg)^{-2a_{\Gamma}(\mu_{h,up},\mu)}H_{up}(\mu_{h,up}),\\
H_{dn}(\mu)&=& \mathrm{exp}\bigl[ 2S(\mu_{h,dn},\mu)-2a_{dn}^V(\mu_{h,dn},\mu)\bigr]\bigg( \frac{-\hat{t}+m_t^2}{\mu_{h,dn}m_t} \bigg)^{-2a_{\Gamma}(\mu_{h,dn},\mu)}H_{dn}(\mu_{h,dn}),
\end{eqnarray}
where $S(\mu_{h,up},\mu)$ and $a_{up}^V$ are defined as~\cite{Becher:2006mr}
\begin{eqnarray}
S(\mu_{h,up},\mu)&=&-\int_{\alpha_s(\mu_{h,up})}^{\alpha_s(\mu)}d\alpha \frac{\Gamma_{\rm cusp}(\alpha)}{\beta(\alpha)}
 \int_{\alpha_s(\mu_{h,up})}^{\alpha}\frac{d\alpha^{\prime}}{\beta(\alpha^{\prime})},\\
a_{up}^V(\mu_{h,up},\mu)&=&-\int_{\alpha_s(\mu_{h,up})}^{\alpha_s(\mu)}d\alpha
\frac{\gamma_{dn}^V(\alpha)}{\beta(\alpha)},
\end{eqnarray}
and similarly for $S(\mu_{h,dn},\mu)$, $a_{\Gamma}$ and $a_{dn}^V$.

\subsection{Jet function}
The jet function $J(p^2,\mu)$ is defined as~\cite{Bauer:2008jx}
\begin{equation}
\theta(p^0) p^- J(p^2,\mu)=\frac{1}{8\pi N_c}
\int \frac{d^4p'}{(2\pi)^4} {\rm Tr}
\la 0|  \bar{\chi}(-p') \bnslash_1
\chi (-p) |0\ra.
\end{equation}
The RG evolution of the jet function is given
by~\cite{Becher:2006mr}
\begin{equation}
 \frac{dJ(p^2,\mu)}{d\ln\mu} = \left( -2 \Gamma_{\rm cusp}
\ln\frac{p^2}{\mu^2} - 2 \gamma^J \right)J(p^2,\mu)
+2\Gcusp\int^{p^2}_0
dq^2\,\frac{J(p^2,\mu)-J(q^2,\mu)}{p^2-q^2}.
\end{equation}
To solve this integro-differential evolution equation, we use the
Laplace transformed jet function:
\begin{equation}
 \widetilde{j}(\ln\frac{Q^2}{\mu^2},\mu)=\int^\infty_0
dp^2\,\exp(-\frac{p^2}{Q^2e^{\gamma_E}}) J(p^2,\mu),
\end{equation}
which satisfies the the RG equation
\begin{equation}
 \frac{d}{d\ln\mu}\widetilde{j}(\ln\frac{Q^2}{\mu^2},\mu)=\left(-2
\Gamma_{\rm cusp}
\ln\frac{Q^2}{\mu^2}-2\gamma^J\right)\widetilde{j}(\ln\frac{Q^2}{\mu^2},\mu).
\end{equation}
Then the jet function at an arbitrary scale $\mu$ is given by
\begin{equation}
 {J}(p^2,\mu)=\exp \bigl[ -4S(\mu_j,\mu)+2a^J(\mu_j,\mu)
\bigr] \widetilde{j}(\partial_{\eta_j}, \mu_j )  \frac{1}{p^2} \left(
\frac{p^2}{\mu^2_j}\right)^{\eta_j}
\frac{e^{-\gamma_E
\eta_j}}{\Gamma(\eta_j)},\label{jet function}
\end{equation}
where $\eta_j=2 a_\Gamma(\muj,\mu)$. The Laplace transformed jet
function $\widetilde{j}(L,\mu)$ at NLO is
\begin{eqnarray}
\widetilde{j}(L,\mu)&=&1+\frac{\alpha_s}{4\pi}\biggl\{ \frac{\Gamma_0^2}{2}L^2+\gamma^J_0L+c^J_1\biggr\},
\end{eqnarray}
where $c_1^J =\left( 7- \frac{2}{3}\pi^2 \right)C_F$.

\subsection{Soft function}
The soft function $S(k^+,\mu)$, which describe soft interactions between all colored particles, is defined as~\cite{Bauer:2008jx}
\begin{eqnarray}
S(k^+,\mu) &=& \frac{1}{N^2_c} \int dk^+ \frac{d^4
k'_s}{(2\pi)^4}\frac{d^4 k_s}{(2\pi)^4} \la 0|
{\mathcal{O}^{\dagger,fedc}_{S}}(k'_s) \delta[k^+ -n_1 \mcdot k_s]
\mathcal{O}^{cdef}_{S} (k_s)| 0 \ra,
\end{eqnarray}
where
\begin{eqnarray}
 \cO_{S}^{cdef}(k_s)&=&\int\,d^4x
e^{-ik_s\,\mcdot \,x} \mathbf{T}\left[ \left( Y^\dagger_{n_b} (x)
Y_{n_a}(x) \right)^{cd}
\left((\tilde{Y}^\dagger_{v_2}(x)\tilde{Y}_{n_1}
(x)\right)^{ef}\right].
\end{eqnarray}
Here $\mathbf{T}$ is the time-ordering operator required to ensure
the proper ordering of soft gluon fields in the soft Wilson
line~\cite{Chay:2004zn}, which is defined as
\begin{equation}
 Y_n(x) = \mathbf{P} \exp\left( ig_s\int^0_{-\infty}ds\,
n\mcdot A^a_s(x+sn)t^a\right)
\end{equation}
for incoming Wilson lines, and
\begin{equation}
 \tilde{Y}_n(x) = \mathbf{P} \exp\left(
-ig_s\int^\infty_{0}ds\,
n\mcdot A^a_s(x+sn)t^a\right)
\end{equation}
for out going Wilson lines, respectively.

\begin{figure}\centering
  \includegraphics[width=0.8\linewidth]{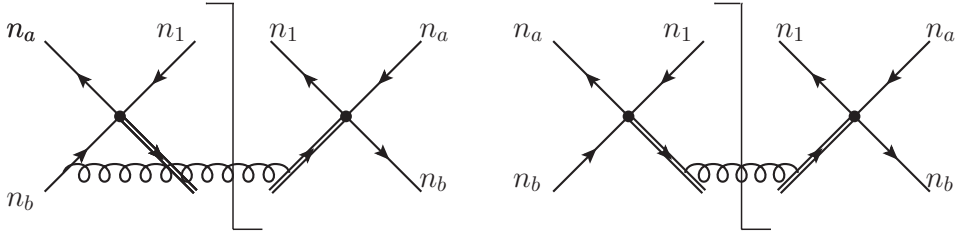}\\
  \caption{Non-vanishing diagrams contributing to the soft function at NLO.
  The contribution from the left and right diagrams are denoted as $S_{bt}$ and $S_{tt}$, respectively.}
  \label{soft function}
\end{figure}

The soft function can be calculated in SCET or in the full theory in
the Eikonal approximation~\cite{Becher:2009th}. In DR, actually, we
only need to calculate the non-vanishing real emission diagrams, as
shown in figure~\ref{soft function},
which give
\begin{eqnarray}\label{eqs:soft}
S^{(1)}_{bt}(k,\mu)&=&\frac{2g_s^2C_F\mu^{2\epsilon}}{(2\pi)^{d-1}}\half
\int_0^{\infty}\hspace{-0.2cm}\mathrm{d}q^+\hspace{-0.1cm}
\int_0^{\infty}\hspace{-0.2cm}\mathrm{d}q^-\hspace{-0.1cm}\int\mathrm{d}^{d-2}q_{\bot}\nn\\
&&\delta(q^+ q^--q_{\bot}^2)\delta(k-n_1\cdot q)\frac{n_b\cdot
v}{(q\cdot n_b)(q\cdot v)},
\end{eqnarray}
and
\begin{eqnarray}
S^{(1)}_{tt}(k,\mu)&=&\frac{-g_s^2C_F\mu^{2\epsilon}}{(2\pi)^{d-1}}\half
\int_0^{\infty}\hspace{-0.2cm}\mathrm{d}q^+\hspace{-0.1cm}
\int_0^{\infty}\hspace{-0.2cm}\mathrm{d}q^-\hspace{-0.1cm}\int\mathrm{d}^{d-2}q_{\bot}\nn\\
&&\delta(q^+ q^--q_{\bot}^2)\delta(k-n_1\cdot q)\frac{1}{(q\cdot
v)^2},
\end{eqnarray}
respectively. Evaluating these integrals, we get
\begin{eqnarray}
S_{bt}(k,\mu)&=&\delta(k)+\frac{2C_F\alpha_s}{4\pi}\biggl\{
 4\biggl[ \frac{\ln \frac{k}{\tilde{\mu}}}{k} \biggr]_{\star}^{[k,\tilde{\mu}]}+\delta(k)c_{bt}^S \biggr\},
 \end{eqnarray}
and
\begin{eqnarray}
S_{tt}(k,\mu)&=&\delta(k)-\frac{C_F\alpha_s}{4\pi}\biggl\{ 4\biggl[
\frac{1}{k} \biggr]_{\star}^{[k,\tilde{\mu}]}+\delta(k)c_{tt}^S
\biggr\},
\end{eqnarray}
respectively, where
$\tilde{\mu}=\mu/\sqrt{(2n_{b\bar{b}})/n_1^{+2}}=(\mu(-\hat{u})m_t)/(2(-\hat{t}+m_t^2)E_1)$.
The explicit expressions of $c_{bt}^S$ and $c_{tt}^S$ are given in
appendix B. And the soft function
$S(k,\mu)=S_{bt}(k,\mu)+S_{bt}(k,\mu)$, similar to the jet function,
satisfies the RG equation
\begin{eqnarray}
\frac{d }{d \ln \mu}S(k,\mu)=\biggl[ -2\Gamma_{\rm cusp}\ln\frac{k}{\tilde{\mu}}+2\gamma^S \biggr]S(k,\mu)
+2\Gamma_{\rm cusp}\int_0^k d k^{\prime}\frac{S(k,\mu)-S(k^{\prime},\mu)}{k-k^{\prime}}.
\end{eqnarray}
The solution to this equation is
\begin{eqnarray}
S(k,\mu)=\exp \bigl[ -2S(\mu_s,\mu)-2a^S(\mu_s,\mu)
\bigr]\widetilde{s} (\partial_{\eta_s},\mu_s)\frac{1}{k}\biggl(
\frac{k}{\tilde{\mu}_s} \biggr)^{\eta_s}
\frac{e^{-\gamma_E\eta_s}}{\Gamma(\eta_s)},
\end{eqnarray}
where $\eta_s=2a_{\Gamma}(\mu_s,\mu)$. The Laplace transformed soft
function $\widetilde{s}(L,\mu)$ at NLO is given by
\begin{eqnarray}
\widetilde{s}(L,\mu)&=&1+\frac{\alpha_s}{4\pi}\biggl\{ \Gamma_0 L^2-2\gamma^S_0 L+ c^S_1 \biggr\}, 
\end{eqnarray}
where $c_1^S=(2c^S_{bt}-c^S_{tt}+\frac{2\pi^2}{3})C_F$.

\subsection{Final resummed differential cross section}
After combining the hard, jet and soft function together, according
to eq.~(\ref{main2}), we obtain the resummed differential cross
section for t-channel single top production
\begin{eqnarray}
\frac{d\hat{\sigma}^{\rm thres}}{d\hat{t}d\hat{u}} &=&
\sum_{ij}\frac{\lambda_{0,ij}}{64\pi N^2_c\hat{s}^2} \nn\\
&& \exp\bigl[ 4S(\mu_{h,up},\mu_{F,up})-2a_{up}^V(\mu_{h,up},\mu_{F,up})\bigr]
  \bigg( \frac{-\hat{t}}{\mu_{h,up}^2} \bigg)^{-2a_{\Gamma}(\mu_{h,up},\mu_{F,up})}H_{up}(\mu_{h,up}) \nn\\
&& \exp\bigl[ 2S(\mu_{h,dn},\mu_{F,dn})-2a_{dn}^V(\mu_{h,dn},\mu_{F,dn})\bigr]
  \bigg( \frac{-\hat{t}+m_t^2}{\mu_{h,dn}m_t} \bigg)^{-2a_{\Gamma}(\mu_{h,dn},\mu_{F,dn})}H_{dn}(\mu_{h,dn})\nn\\
&& \exp \bigl[ -4S(\mu_j,\mu_{F,up})+2a^J(\mu_j,\mu_{F,up})\bigr] \left( \frac{m_t^2}{\mu^2_j} \right)^{\eta_j} \nn\\
&& \exp \bigl[ -2S(\mu_s,\mu_{F,dn})-2a^S(\mu_s,\mu_{F,dn}) \bigr]\biggl( \frac{m_t(-\hat{t}+m_t^2)}{\mu_s(-\hat{u}) } \biggr)^{\eta_s}  \nn\\
&& \widetilde{j}(\partial_{\eta}+L_j , \mu_j )\widetilde{s}(\partial_{\eta}+L_s,\mu_s)
 \frac{1}{s_4}\left( \frac{s_4}{m_t^2} \right)^{\eta}
 \frac{e^{-\gamma_E\eta}}{\Gamma(\eta)},
\end{eqnarray}
where $\eta=\eta_j+\eta_s$ and $L_j=\ln (m_t^2/\mu^2_j), L_s=\ln (m_t(-\hat{t}+m_t^2))/(\mu_s(-\hat{u}))$.
In the above expression, we have chosen
$\mu=\mu_{F,up}$ or $\mu=\mu_{F,dn}$ to avoid the evolution
of the parton distribution functions.

If we set scales $\mu_{h,up}$, $\mu_{h,dn}$, $\mu_{j}$, $\mu_{s}$
equal to the common scale $\mu$, which is conveniently chosen as the
factorization scale $\mu_{F,up}=\mu_{F,dn}=\mu_{F}$, then we recover
the threshold singular plus distributions, which should appear in
the fixed-order calculation. Up to order $\alpha_s^2$, we have
\begin{eqnarray}
\frac{\lambda_{0,ij}}{64\pi N^2_c\hat{s}^2}\frac{d\hat{\sigma}_{ij}^{\rm thres}}{d\hat{t}d\hat{u}} &=&\delta(s_4)+\frac{\alpha_s}{4\pi}\biggl\{
3\Gamma_0\bigg[\frac{\ln (s_4/m_t^2)}{s_4} \bigg]_{+}+\big[\gamma^J_0-2\gamma^S_0+(L_j+2L_s)\Gamma_0\big]\biggr[\frac{1}{s_4} \biggl]_{+} \biggr\}
\nn\\
&&\hspace{-3.5cm}+\bigg(\frac{\alpha_s}{4\pi}\bigg)^2 \bigg\{ \frac{9\Gamma_0^2}{2}\biggr[\frac{\ln^3 (s_4/m_t^2)}{s_4} \biggl]_{+}+\big[\frac{9}{2}(L_j+2L_s)\Gamma_0^2-\half(5\beta_0-9\gamma^J_0-18\gamma^S_0)\Gamma_0\big]\biggr[\frac{\ln^2 (s_4/m_t^2)}{s_4} \biggl]_{+}
\nn\\
&&\hspace{-3.5cm}+\bigl[\frac{5\Gamma_0^2}{2}L_j^2+(4L_s\Gamma_0^2+(5\gamma^J_0-4\gamma^S_0-\beta_0)\Gamma_0)L_j
+7\Gamma_0^2L_s^2+(4\gamma^J_0-14\gamma^S_0-4\beta_0)L_s-\frac{9\pi^2}{4}\Gamma_0^2
\nn\\
&&\hspace{-3.5cm}+3(c^H_1+c^J_1+c^S_1)\Gamma_0
+(\gamma^J_0-2\gamma^S_0)^2-\beta_0(\gamma^J_0-4\gamma^S_0)+3\Gamma_1 \bigr]\bigg[\frac{\ln (s_4/m_t^2)}{s_4}\bigg]_{+}+\bigl[\frac{\Gamma^2_0}{2}L_j^3+\{L_s\Gamma_0^2
\nn\\
&&\hspace{-3.5cm}+\half(3\gamma^J_0-2\gamma^S_0-\beta_0)\Gamma_0\}L_j^2
+\{\Gamma^2_0L_s^2+2(\gamma^J_0-\gamma^S_0)\Gamma_0L_s+(c^J_1+c^S_1)\Gamma_0
+(\gamma^J_0-2\gamma^S_0-\beta_0)\gamma^J_0
\nn\\
&&\hspace{-3.5cm}-\frac{3\pi^2}{4}\Gamma^2_0+\Gamma_1\}L_j
+2\Gamma_0^2L_s^3+(\gamma^J_0-6\gamma^S_0-2\beta_0)\Gamma_0L_s^2
+\{-\frac{3\pi^2}{2}\Gamma_0^2+2(c^J_1+c^S_1)\Gamma_0
\nn\\
&&\hspace{-3.5cm}+2(2\gamma^S_0-\gamma^J_0+2\beta_0)\gamma^S_0+2\Gamma_1\}L_s+9\zeta_3\Gamma_0^2
-(\frac{3\gamma^J_0}{4}-\frac{3\gamma^S_0}{2}-\frac{5\beta_0}{12})\pi^2\Gamma_0
+(\gamma^J_0-2\gamma^S_0-\beta_0)c^J_1
\nn\\
&&\hspace{-3.5cm}+(\gamma^J_0-2\gamma^S_0-2\beta_0)c^S_1+\gamma^J_1-2\gamma^S_1+\big\{\gamma^J_0-2\gamma^S_0+(L_j+2L_s)\Gamma_0\big\}c^H_1
\bigr]\biggl[\frac{1}{s_4} \biggr]_{+}
\bigg\},
\end{eqnarray}
where $c^H_1=c^{H,up}_1+c^{H,dn}_1$.
We find that all $\mathcal {O}(\alpha_s)$ and two front $\mathcal
{O}(\alpha_s^2)$ singular plus distribution terms coincide with
Kidonakis'~\cite{Kidonakis:2006bu}.

Including the remaining terms in the NLO result which we do not resum,
we obtain the final resummed differential cross section
\begin{eqnarray}
\frac{d\hat{\sigma}^{\rm
RES}}{d\hat{t}d\hat{u}}&=&\frac{d\hat{\sigma}^{\rm
thres}}{d\hat{t}d\hat{u}}\Big
|_{\mu_{F,up},\mu_{F,dn},\mu_{h,up},\mu_{h,dn},\mu_j,\mu_s}\nn\\ &&
+\frac{d\hat{\sigma}^{\rm NLO}}{d\hat{t}d\hat{u}}\Big
|_{\mu_{F,up},\mu_{F,dn}} -\frac{d\hat{\sigma}^{\rm
thres}}{d\hat{t}d\hat{u}}\Big
|_{\mu_{F,up}=\mu_{F,dn}=\mu_{h,up}=\mu_{h,dn}=\mu_j=\mu_s}.
\end{eqnarray}

\section{Numerical Discussion}\label{sec:4}
In this section, we discuss the numerical results of threshold
resummation in t-channel single top production via SCET. In our
calculation, there are four scales, except the two factorization
scales, $\mu_{h,up},\mu_{h,dn},\mu_j,\mu_s$ explicitly,  which are
all arbitrary in principle and our final results should not depend
on them.  However, because the Wilson coefficients in each matching,
expressed as hard, jet and soft functions, respectively, and the
anomalous dimensions are evaluated in fixed-order perturbation
theory, there are residual dependence on these scales. To
illustrate the reliability of our evaluation, first, we investigate
these scale uncertainties. In the discussion below, we focus on the
scenario at the Tevatron and give a cut to
$\sqrt{-\hat{t}}$\footnote{For s-channel processes, the invariant
mass of final state particles provides a natural cut to the transfer
momentum $\sqrt{\hat{s}}$.}, which is the transfer momentum of this process,
because, based on the view point of effective theory, only processes
of large transfer momentum are considered as hard processes with
which we are concerned.
\subsection{Scale choices and uncertainties}
First, we discuss the dependence of $H_{up}(\mu_{h,up})$ on
$\mu_{h,up}$. From eq.(\ref{hard_function}), in order to avoid large logarithms, we choose
$\sqrt{-\hat{t}}$ as our natural hard(up) scale. The left curves in figure~\ref{hrdscl
dependence}~illustrate the RG effects reduce the dependence of
$H_{up}(\mu_{h,up})$ on $\mu_{h,up}$. And the correction induced by
the RG-improved $H_{up}(\mu_{h,up})$ to the LO cross section is about
$-24\%$, which is significant.
\begin{figure}
  \includegraphics[width=0.48\linewidth]{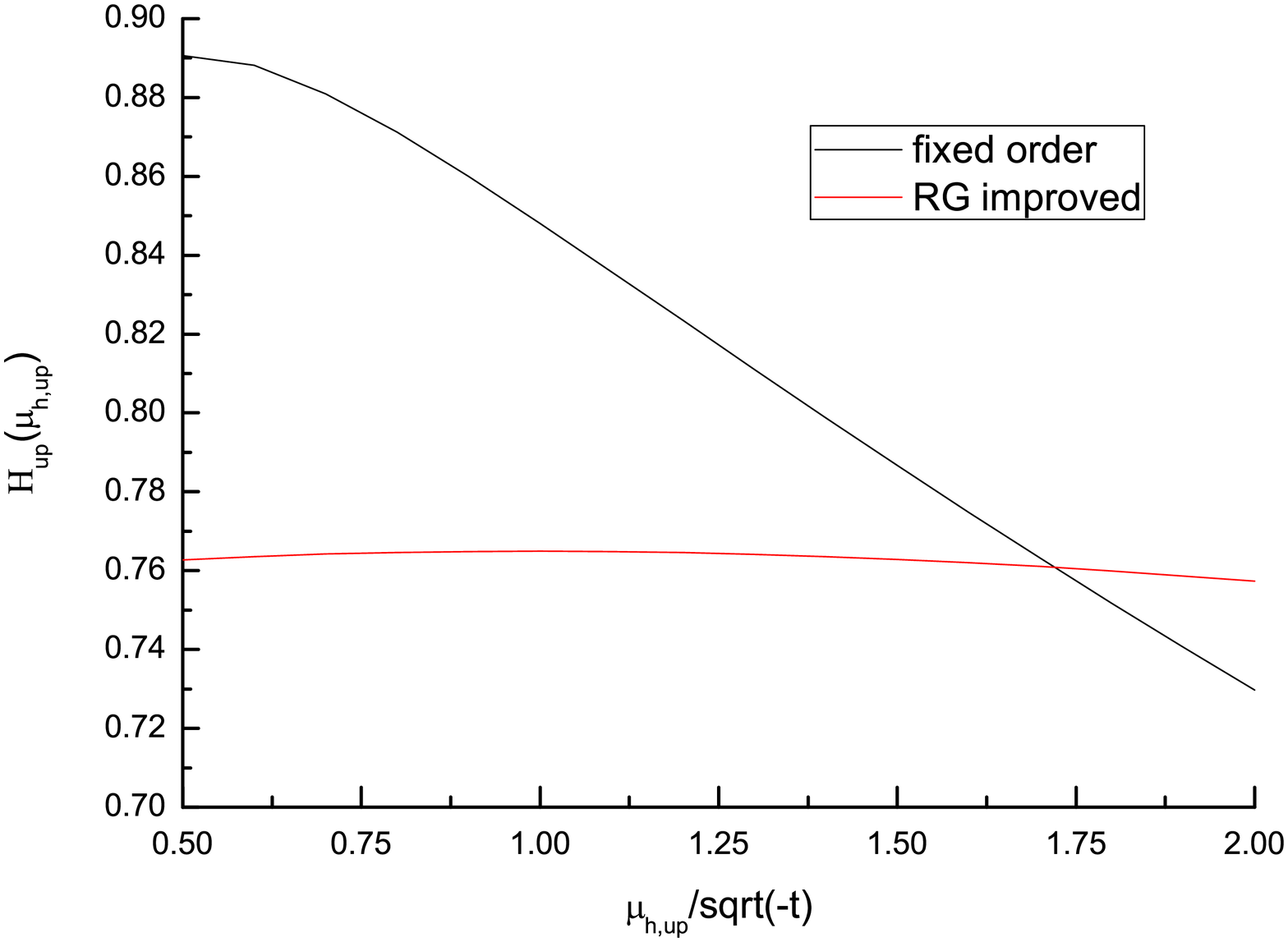}
  \includegraphics[width=0.48\linewidth]{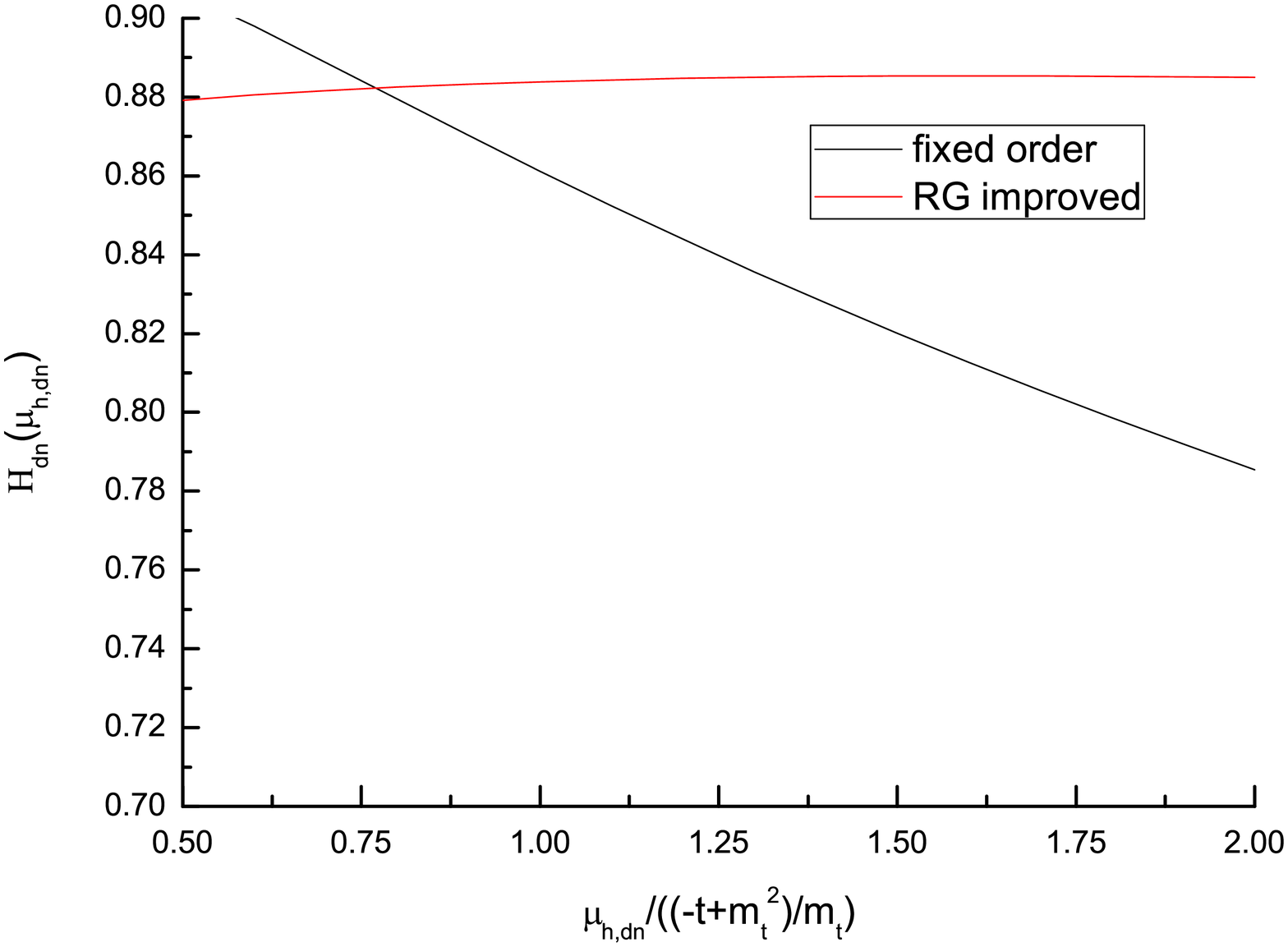}
  \caption{ The variations of $H_{up}(\mu_{h,up})$ and $H_{dn}(\mu_{h,dn})$ with $\mu_{h,up}$ and $\mu_{h,dn}$, respectively.}
  \label{hrdscl dependence}
\end{figure}
The right curves in figure~\ref{hrdscl dependence}~show the RG
effects reduce the dependence of $H_{dn}(\mu_{h,dn})$ on
$\mu_{h,dn}$, where we choose $(-\hat{t}+m_t^2)/m_t$ as our natural hard(down) scale.
From these curves, we can see that the scale
dependence is reduced and its RG-improved correction to LO cross
section is about $-11\%$.

\begin{figure}
  \includegraphics[width=0.48\linewidth]{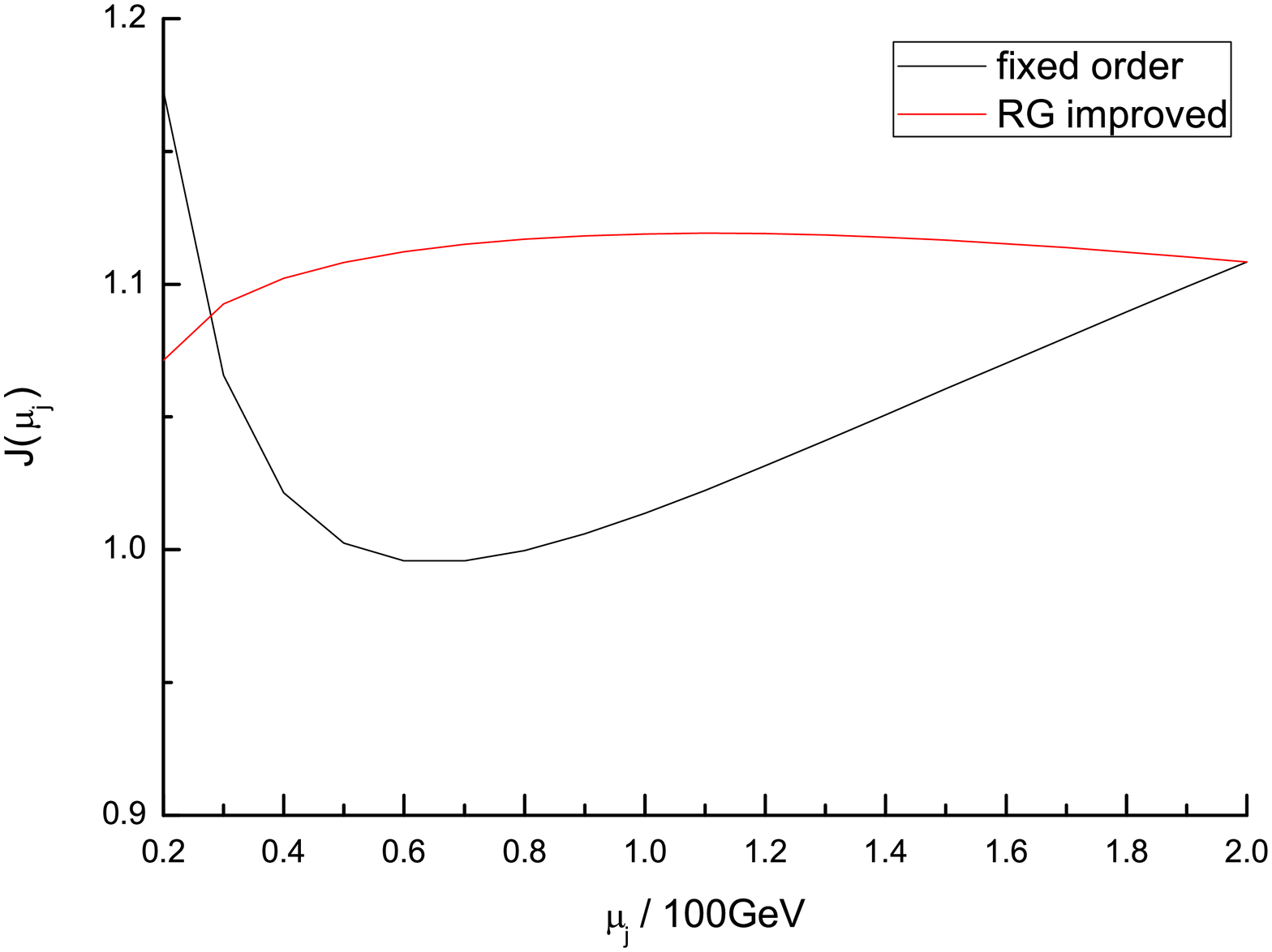}
  \includegraphics[width=0.48\linewidth]{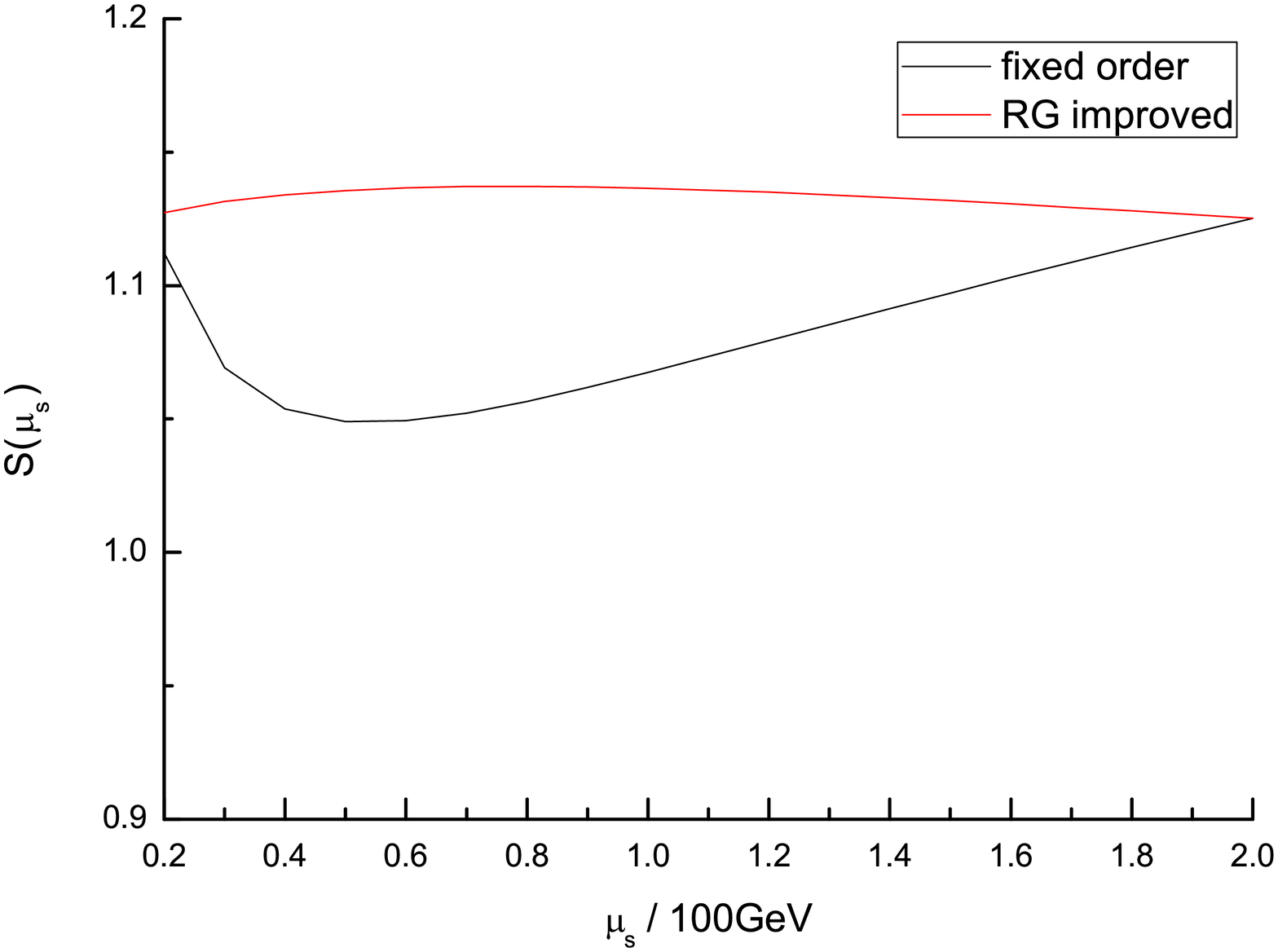}
  \caption{ The variations of $J(\mu_{j})$ and $S(\mu_{s})$ with $\mu_{j}$ and $\mu_{s}$, respectively.}
  \label{jetscl dependence}
\end{figure}

Then, we examine the dependence of jet function on $\mu_j$. Unlike the case of
hard functions, because we perform the integration convoluting the jet and soft
functions analytically, we can only choose the natural jet scale through the
numerical results. In figure~\ref{jetscl dependence}, we show that the
natural jet scale is about $60$~GeV around which the contribution of
the fixed order jet function is minimal. Besides, the RG-improved jet function vary
slowly, which indicates that the scale dependence is significantly
 reduced. The correction induced by the RG-improved jet
function to the LO cross section is about $+12\%$.

The same case happens in the soft function. In principle, we may
consider $2E_1 k (-\hat{t}+m_t^2)/(-\hat{u})/m_t$ as the natural soft
scale. But, in practice, from the numerical results in figure~\ref{jetscl dependence},
we set the natural soft scale at $50$~GeV,
and find that the correction induced by the RG-improved soft function to
the LO cross section is about $+14\%$.

\subsection{Resummed cross sections}
We have chosen all the natural scales needed in this process. Now we
give the numerical results of the resummed cross section. When
discussing each scale dependence, we fixed the other scales at the
natural scales chosen in the last subsection.

\begin{table}[ht]\centering
\begin{tabular}{|c|c|c|c|}
  \hline
   & $\mu_{F,up}$ & $\mu_{F,dn}$ & $\mu_{F,up}$ \& $\mu_{F,dn}$ \\
   \hline
   $\sigma_{\rm LO}$(pb)                   &\quad $0.959^{+0.064}_{-0.055}$ \hspace{0.2cm} &\quad $0.959^{-0.058}_{+0.041}$\hspace{0.2cm} &\quad $0.959^{-0.001}_{+0.016}$\hspace{0.2cm} \\
   \hline
   $\sigma_{\rm NLO}$(pb)                  &\quad $0.977^{+0.006}_{-0.004}$ \hspace{0.2cm} &\quad $0.977^{-0.032}_{+0.021}$\hspace{0.2cm} &\quad $0.977^{-0.030}_{+0.026}$\hspace{0.2cm}  \\
  \hline
  $\sigma_{\rm RES}$(pb)
   &\quad $0.948^{+0.047}_{-0.033}$ \hspace{0.2cm} &\quad $0.948^{-0.041}_{+0.030}$\hspace{0.2cm} &\quad $0.948^{-0.012}_{+0.006}$\hspace{0.2cm} \\
  \hline
\end{tabular}
 \caption{The variations of the resummed cross section and fixed order cross section with factorization scales
 at the Tevatron with $\sqrt{S}$=1.96~TeV,
  choosing $ m_t=175\rm ~GeV$ and transfer momentum cut $\sqrt{-\hat{t}}>80 {\rm ~GeV}$.}
 \label{table-comparation}
\end{table}

In table~\ref{table-comparation}, we vary the factorization scales over the ranges
$100{\rm~GeV}<\mu_{F,up}, \mu_{F,dn}<400{\rm~GeV}$. And table \ref{table-comparation} shows that the
resummation effects decrease the NLO cross section by about $3\%$.
This is reasonable because we have seen above that the virtual
effects are negative, and their absolute values are larger than
those of jet and soft effects. These effects, when resummed to all
order, would contribute large negative corrections which even
overwhelm the positive jet and soft contributions. As a result, the
resummed cross section is less than the LO order if the same parton
distribution functions are applied, just as we did in this work.
From table \ref{table-comparation} we also can see the factorization
scale dependence of resummed cross sections is reduced when the two
factorization scales vary simultaneously, compared with NLO cross
sections. But, if the factorization scales vary separately, the
scale dependence get worse, thought
 still better than that of LO cross sections.

In table \ref{table-scale}, we show scale dependencies of
the resummed cross section, where the scales vary over the ranges
$\sqrt{-\hat{t}}/2<\mu_{h,up}<2\sqrt{-\hat{t}}$,
$(-\hat{t}+m_t^2)/m_t/2<\mu_{h,dn}<2(-\hat{t}+m_t^2)/m_t$,
$30{\rm~GeV}<\mu_{j}<120{\rm~GeV}$ and
${\rm25~GeV}<\mu_{s}<{\rm100~GeV}$. And we can see that their
uncertainties are all very small.

\begin{table}[ht]\centering
\begin{tabular}{|c|c|c|c|c|}
  \hline
   & $\mu_{h,up}$ & $\mu_{h,dn}$ & $\mu_{j}$ & $\mu_{s}$ \\
   \hline
  $\sigma_{\rm RES}$(pb)
  &\quad $0.948^{+0.001}_{-0.010}$ \hspace{0.2cm} &\quad $0.948^{-0.001}_{+0.005}$\hspace{0.2cm} &\quad $0.948^{-0.003}_{+0.009}$\hspace{0.2cm}&\quad $0.948^{-0.001}_{+0.005}$\hspace{0.2cm} \\
  \hline
\end{tabular}
 \caption{The~$\mu_{h,up}$, $\mu_{h,dn}$, $\mu_{j}$ and $\mu_{s}$ scale dependencies of resummed cross section
 at the Tevatron with $\sqrt{S}$=1.96~TeV,
taking $ m_t=175\rm ~GeV$ and transfer momentum cut
$\sqrt{-\hat{t}}>80 {\rm ~GeV}$.}
 \label{table-scale}
\end{table}

In table \ref{table-mt}, we show how the value of $m_t$ affects our
results. When the value of $m_t$ varies from $171$~GeV to $175$~GeV,
the resummed cross sections vary by about $6\%$.

\begin{table}[ht]\centering
\begin{tabular}{|c|c|c|c|c|c|}
  \hline
   $m_{t}$(GeV)& $171$ & 172 & 173 & 174 & 175 \\
   \hline
   $\sigma_{\rm LO}$(pb)  &\quad 1.032 \hspace{0.2cm} &\quad 1.013 \hspace{0.2cm} &\quad 0.995 \hspace{0.2cm} &\quad 0.977 \hspace{0.2cm} &\quad 0.959 \hspace{0.2cm} \\
   \hline
   $\sigma_{\rm NLO}$(pb) &\quad 1.037 \hspace{0.2cm} &\quad 1.026 \hspace{0.2cm} &\quad 1.010 \hspace{0.2cm} &\quad 0.987 \hspace{0.2cm} &\quad 0.977 \hspace{0.2cm} \\
  \hline
  $\sigma_{\rm RES}$(pb)
  &\quad 1.008 \hspace{0.2cm} &\quad 0.997 \hspace{0.2cm} &\quad 0.982 \hspace{0.2cm} &\quad 0.959 \hspace{0.2cm} &\quad 0.948 \hspace{0.2cm} \\
  \hline
\end{tabular}
 \caption{The $m_t$ dependence of resummed cross section at the Tevatron with
 $\sqrt{S}$=1.96~TeV,
taking transfer momentum cut $\sqrt{-\hat{t}}>80 {\rm ~GeV}$.}
\label{table-mt}
\end{table}

Table~\ref{table-cut} gives the transfer momentum cut dependence of
the cross sections. It shows that the resummed cross section gets
smaller when the transfer momentum cut is decreased since choosing a
smaller transfer momentum cut means that more hard effects are
resummed. But as mentioned above, the transfer momentum cut can not
be chosen too small for a hard process. Therefore, we choose
$80$~GeV as the natural transfer momentum cut in the numerical
calculations.
\begin{table}[ht]\centering
\begin{tabular}{|c|c|c|c|c|c|}
  \hline
   $\sqrt{-\hat{t}}$(GeV)& $>60$ & $>70$ & $>80$ & $>90$ & $>100$ \\
   \hline
  $\sigma_{\rm RES}$(pb)
  &\quad 0.920 \hspace{0.2cm} &\quad 0.936 \hspace{0.2cm} &\quad 0.948 \hspace{0.2cm} &\quad 0.957 \hspace{0.2cm} &\quad 0.964 \hspace{0.2cm} \\
  \hline
\end{tabular}
 \caption{The transfer momentum cut dependence of resummed cross section at the Tevatron with
 $\sqrt{S}$=1.96~TeV,
taking $ m_t=175\rm ~GeV$.}
 \label{table-cut}
\end{table}

 In tables~\ref{table-topLHC7}, \ref{table-topLHC10},
 \ref{table-topLHC14}, \ref{table-antitopLHC7}, \ref{table-antitopLHC10}, and \ref{table-antitopLHC14},
we present the results for the single top (antitop) production at the
LHC for different top quark mass with $\sqrt{S}=7,10,{\rm and}
~14$~TeV, respectively. We can see that the resummation effects
decrease the NLO cross sections by about $2\%$ when the transfer
momentum cut is chosen as $80$~GeV. And the factorization scale
dependencies of the cross section are reduced also.
\begin{table}[hbt]\centering
\begin{tabular}{|c|c|c|c|c|c|}
  \hline
   $m_{t}$(GeV)& $171$ & 172 & 173 & 174 & 175 \\
   \hline
   $\sigma_{\rm LO}$(pb)  & $44.9^{-3.1}_{+2.2}$  & $44.4^{-3.1}_{+2.1}$  & $43.9^{-3.0}_{+2.1}$
                          & $43.5^{-3.0}_{+2.1}$  & $43.0^{-2.9}_{+2.0}$  \\
   \hline
   $\sigma_{\rm NLO}$(pb) & $42.6^{-0.8}_{+1.0}$  & $42.2^{-0.7}_{+1.2}$  & $41.9^{-0.6}_{+0.9}$
                          & $41.6^{-0.8}_{+0.8}$  & $41.1^{-0.7}_{+0.9}$  \\
  \hline
  $\sigma_{\rm RES}$(pb)
  & $41.7^{-0.1}_{+0.2}$  & $41.3^{-0.1}_{+0.3}$  & $40.9^{-0.1}_{+0.1}$
  & $40.7^{-0.1}_{+0.1}$  & $40.2^{-0.1}_{+0.1}$  \\
  \hline
\end{tabular}
 \caption{The cross sections for t-channel single top production at LHC with
 $\sqrt{S}$=7~TeV,
 choosing transfer momentum cut $\sqrt{-\hat{t}}>80 {\rm
~GeV}$. The factorization scale uncertainties are also shown.}
 \label{table-topLHC7}
 \begin{tabular}{|c|c|c|c|c|c|}
  \hline
   $m_{t}$(GeV)& $171$ & 172 & 173 & 174 & 175 \\
   \hline
   $\sigma_{\rm LO}$(pb)  & $90.9^{-7.7}_{+5.9}$  & $90.0^{-7.6}_{+5.8}$  & $89.2^{-7.5}_{+5.7}$
                          & $88.4^{-7.4}_{+5.6}$  & $87.5^{-7.4}_{+5.5}$  \\
   \hline
   $\sigma_{\rm NLO}$(pb) & $86.1^{-1.9}_{+2.1}$  & $85.5^{-1.9}_{+1.5}$  & $84.6^{-1.5}_{+1.8}$
                          & $83.3^{-0.8}_{+2.5}$  & $83.0^{-1.2}_{+1.6}$  \\
  \hline
  $\sigma_{\rm RES}$(pb)
  & $84.4^{-0.5}_{+0.4}$  & $83.8^{-0.6}_{+0.1}$  & $82.9^{-0.1}_{+0.2}$
  & $81.6^{-0.1}_{+0.8}$  & $81.3^{-0.1}_{+0.6}$  \\
  \hline
\end{tabular}
 \caption{The cross sections for t-channel single top production at LHC with
 $\sqrt{S}$=10~TeV,
choosing transfer momentum cut $\sqrt{-\hat{t}}>80 {\rm ~GeV}$. The
factorization scale uncertainties are also shown.}
 \label{table-topLHC10}
 \begin{tabular}{|c|c|c|c|c|c|}
  \hline
   $m_{t}$(GeV)& $171$ & 172 & 173 & 174 & 175 \\
   \hline
   $\sigma_{\rm LO}$(pb)  & $167.0^{-16.5}_{+13.2}$  & $165.6^{-16.3}_{+13.0}$  & $164.2^{-16.1}_{+12.9}$
                          & $162.7^{-16.0}_{+12.7}$  & $161.3^{-15.8}_{+12.8}$  \\
   \hline
   $\sigma_{\rm NLO}$(pb) & $157.2^{-3.4}_{+3.8}$  & $156.9^{-3.6}_{+1.8}$ & $155.3^{-3.3}_{+3.3}$
                          & $154.9^{-4.8}_{+2.9}$  & $152.5^{-3.0}_{+4.5}$ \\
  \hline
  $\sigma_{\rm RES}$(pb)
  & $154.3^{-1.1}_{+1.5}$  & $154.0^{-1.2}_{+0.1}$  & $152.4^{-1.0}_{+0.4}$
  & $152.0^{-2.4}_{+0.1}$  & $150.0^{-0.7}_{+1.5}$  \\
  \hline
\end{tabular}
 \caption{The cross sections for t-channel single top production at LHC with
 $\sqrt{S}$=14~TeV,
choosing transfer momentum cut $\sqrt{-\hat{t}}>80 {\rm ~GeV}$. The
factorization scale uncertainties are also shown.}
 \label{table-topLHC14}
\end{table}

\begin{table}[hbt]\centering
\begin{tabular}{|c|c|c|c|c|c|}
  \hline
   $m_{t}$(GeV)& $171$ & 172 & 173 & 174 & 175 \\
   \hline
   $\sigma_{\rm LO}$(pb)  & $23.9^{-1.7}_{+1.2}$  & $23.6^{-1.6}_{+1.1}$  & $23.4^{-1.6}_{+1.1}$
                          & $23.1^{-1.6}_{+1.1}$  & $22.9^{-1.6}_{+1.1}$  \\
   \hline
   $\sigma_{\rm NLO}$(pb) & $22.7^{-0.3}_{+0.8}$  & $22.6^{-0.6}_{+0.4}$ & $22.4^{-0.5}_{+0.4}$
                          & $22.1^{-0.5}_{+0.5}$  & $21.9^{-0.4}_{+0.6}$ \\
  \hline
  $\sigma_{\rm RES}$(pb)
  & $22.3^{-0.1}_{+0.5}$  & $22.2^{-0.4}_{+0.1}$  & $22.1^{-0.2}_{+0.1}$
  & $21.8^{-0.2}_{+0.2}$  & $21.5^{-0.1}_{+0.3}$  \\
  \hline
\end{tabular}
 \caption{The cross sections for t-channel single antitop production at LHC with
 $\sqrt{S}$=7~TeV,
choosing transfer momentum cut $\sqrt{-\hat{t}}>80 {\rm ~GeV}$. The
factorization scale uncertainties are also shown.}
 \label{table-antitopLHC7}
 \begin{tabular}{|c|c|c|c|c|c|}
  \hline
   $m_{t}$(GeV)& $171$ & 172 & 173 & 174 & 175 \\
   \hline
   $\sigma_{\rm LO}$(pb)  & $52.5^{-4.5}_{+3.4}$  & $52.0^{-4.5}_{+3.4}$  & $51.5^{-4.4}_{+3.3}$
                          & $51.0^{-4.4}_{+3.3}$  & $50.5^{-4.3}_{+3.3}$  \\
   \hline
   $\sigma_{\rm NLO}$(pb) & $49.6^{-1.0}_{+1.2}$  & $49.2^{-1.3}_{+0.9}$  & $48.7^{-1.1}_{+1.1}$
                          & $48.2^{-0.9}_{+1.2}$  & $47.8^{-1.1}_{+1.1}$  \\
  \hline
  $\sigma_{\rm RES}$(pb)
  & $48.9^{-0.4}_{+0.4}$ & $48.5^{-0.7}_{+0.2}$  & $48.0^{-0.5}_{+0.4}$
  & $47.5^{-0.3}_{+0.5}$ & $47.1^{-0.5}_{+0.4}$  \\
  \hline
\end{tabular}
 \caption{The cross sections for t-channel single antitop production at LHC with
 $\sqrt{S}$=10~TeV,
choosing transfer momentum cut $\sqrt{-\hat{t}}>80 {\rm ~GeV}$. The
factorization scale uncertainties are also shown.}
 \label{table-antitopLHC10}
 \begin{tabular}{|c|c|c|c|c|c|}
  \hline
   $m_{t}$(GeV)& $171$ & 172 & 173 & 174 & 175 \\
   \hline
   $\sigma_{\rm LO}$(pb)  & $103.1^{-10.4}_{+8.3}$  & $102.2^{-10.3}_{+8.2}$  & $101.2^{-10.2}_{+8.1}$
                          & $100.3^{-10.1}_{+8.0}$  & $99.4^{-10.0}_{+7.9}$  \\
   \hline
   $\sigma_{\rm NLO}$(pb) & $96.7^{-2.0}_{+1.9}$  & $96.3^{-2.6}_{+1.4}$  & $95.1^{-2.2}_{+1.8}$
                          & $94.1^{-2.5}_{+2.1}$  & $93.1^{-2.5}_{+1.8}$ \\
  \hline
  $\sigma_{\rm RES}$(pb)
  & $95.4^{-0.9}_{+0.4}$  & $95.0^{-1.4}_{+0.1}$ & $93.8^{-1.0}_{+0.6}$
  & $92.8^{-1.3}_{+0.6}$  & $91.8^{-1.4}_{+0.8}$ \\
  \hline
\end{tabular}
 \caption{The cross sections for t-channel single antitop production at LHC with
 $\sqrt{S}$=14~TeV,
choosing transfer momentum cut $\sqrt{-\hat{t}}>80 {\rm ~GeV}$. The
factorization scale uncertainties are also shown.}
 \label{table-antitopLHC14}
\end{table}

\subsection{Combined s and t channel cross sections}
Table~\ref{table-exp} shows the combined numerical results for
s-~\cite{Zhu:2010mr} and t-channel single top production at the
Tevatron. From table~\ref{table-exp}, we see that our result is
closer to the experimental result~\cite{Group:2009qk} than the one
reported in the previous
literatures\cite{Kidonakis:2006bu,Kidonakis:2007ej,Kidonakis:2010tc}.

\begin{table}[ht]\centering
\begin{tabular}{ccccc}
  \hline\hline
   & NLO\cite{Harris:2002md,Sullivan:2004ie,
Campbell:2004ch} & Res.\cite{Kidonakis:2006bu,Kidonakis:2007ej,Kidonakis:2010tc} & Res.in SCET & Experiments.\cite{Group:2009qk}   \\
   \hline
   \hline
    s-channel  & 0.99 pb  & 1.12 pb  & 1.04 pb  &\quad ---   \\
   \hline
    t-channel  & 2.15 pb  & 2.34 pb  & 2.04 pb  &\quad ---  \\
  \hline
   combined s- and t-channel
  & 3.14 pb  & 3.46 pb  & 3.08 pb  & 2.76 pb   \\
  \hline\hline
\end{tabular}
 \caption{Combination of s- and t-channel single top production at the Tevatron with $\sqrt{S}$=1.96~TeV.
}
  \label{table-exp}
\end{table}

\section{Conclusion}\label{sec:5}
We have studied the factorization and resummation of t-channel
single top (antitop) quark production in the Standard Model at both
the Tevatron and the LHC with SCET. Our results show that the
resummation effects decrease the NLO cross sections by about $3\%$
at the Tevatron and about $2\%$ at the LHC, respectively. And the
resummation effects significantly reduce the factorization scale
dependence of the total cross section when the two factorization
scales vary simultaneously, compared with the NLO results. We also
show that when our numerical results for s-~\cite{Zhu:2010mr} and
t-channel single top production at the Tevatron are combined, it is
closer to the experimental result~\cite{Group:2009qk} than the one
reported in the previous
literatures\cite{Kidonakis:2006bu,Kidonakis:2007ej,Kidonakis:2010tc}
.

\acknowledgments
This work was supported in part by the National Natural
Science Foundation of China, under Grants No.~10721063,
No.~10975004 and No.~10635030.

\appendix
\section{Relevant anomalous dimensions and matching
coefficients}
The various anomalous dimensions
needed in our calculations can be found, e.g., in~\cite{Becher:2006mr,Becher:2007ty,Becher:2009th}. We
list them below for convenience of the reader. The QCD
$\beta$
function is
\begin{equation}
 \beta (\alpha_s) = -2 \alpha_s \left[ \beta_0
\frac{\alpha_s}{4\pi} + \beta_1 \left(
\frac{\alpha_s}{4\pi} \right)^2 + \cdots \right],
\end{equation}
with expansion coefficients
\begin{eqnarray}
 \beta_0 &=& \frac{11}{3} C_A - \frac{4}{3} T_F n_f,
\nn
\\
\beta_1 &=& \frac{34}{3}C^2_A - \frac{20}{3} C_A T_F n_f -
4C_F T_F n_f,
\nn
\\
\beta_2 &=& \frac{2857}{54} C^3_A + \left( 2 C^2_F -
\frac{205}{9} C_F C_A - \frac{1415}{27} C^2_A \right) T_F
n_f + \left( \frac{44}{9}C_F + \frac{158}{27} C_A \right)
T^2_F n^2_f,
\end{eqnarray}
where $C_A=3$, $T_F=1/2$ for QCD, and $n_f$ is
the number of active quark flavor.

The cusp anomalous dimension is
\begin{equation}
\label{cuspa}
 \Gamma_{\rm cusp} (\alpha_s) = \Gamma_0\frac{\alpha_s}{4\pi} +
\Gamma_1 \left(
\frac{\alpha_s}{4\pi} \right)^2 + \cdots,
\end{equation}
with
\begin{eqnarray}
 \Gamma_0 &=& 4C_F,
\nn
\\
\Gamma_1 &=& 4C_F \left[ \left( \frac{67}{9} -
\frac{\pi^2}{3} \right) C_A - \frac{20}{9} T_F n_f \right],
\nn
\\
\Gamma_2 &=& 4C_F \left[ C^2_A \left(
\frac{245}{6} -
\frac{134}{27}\pi^2 + \frac{11}{45}\pi^4 +
\frac{22}{3}\zeta_3 \right) + C_A T_F n_f \left(
-\frac{418}{27} + \frac{40}{27}\pi^2 - \frac{56}{3}\zeta_3
\right)
\right.
\nn
\\
&&\left. + C_F T_F n_f \left( -\frac{55}{3} + 16 \zeta_3
\right) - \frac{16}{27} T^2_F n^2_f \right].
\end{eqnarray}

The other anomalous dimensions are expanded as
eq.~(\ref{cuspa}), and their expansion coefficients are
\begin{eqnarray}
 \gamma^{0}_q &=& -3C_F,
\nn
\\
\gamma^{1}_q &=& C^2_F\left(-\frac{3}{2}+2\pi^2-24
\zeta_3\right) + C_F
C_A \left( -\frac{961}{54}-\frac{11}{6}\pi^2 + 26\zeta_3
\right) + C_F T_F n_f \left( \frac{130}{27} +
\frac{2}{3}\pi^2 \right),
\nn
\\
\gamma^{0}_Q &=& -2C_F,
\nn
\\
\gamma^{1}_Q &=& C_F C_A \left( \frac{2}{3}\pi^2 -
\frac{98}{9} - 4 \zeta_3 \right) + \frac{40}{9} C_F T_F n_f,
\nn
\\
\gamma^{0}_\phi &=& 3C_F,
\nn
\\
\gamma^{1}_\phi &=& C^2_F \left( \frac{3}{2} - 2 \pi^2 +
24 \zeta_3 \right) + C_F C_A \left( \frac{17}{6} +
\frac{22}{9} \pi^2 - 12 \zeta_3 \right) - C_F T_F n_f
\left( \frac{2}{3} + \frac{8}{9} \pi^2 \right),
\nn
\\
\gamma^{0}_j &=& -3C_F,
\nn
\\
\gamma^{1}_j &=& C^2_F \left( -\frac{3}{2} + 2 \pi^2-
24 \zeta_3 \right) + C_F C_A \left( -\frac{1769}{54} -
\frac{11}{9} \pi^2 + 40 \zeta_3 \right)
\nn
\\
&& + C_F T_F n_f
\left( \frac{242}{27} + \frac{4}{9} \pi^2 \right).
\end{eqnarray}
$\gamma^V_{up}$, $\gamma^V_{dn}$ and $\gamma^S$ can be obtained from the
anomalous dimensions above through the following equations:
\begin{eqnarray}
  \gamma^V_{up} &=& 2 \gamma_q ,\nn\\
  \gamma^V_{dn} &=& \gamma_q + \gamma_Q,\nn\\
\gamma^S &=& -2 \gamma_\phi - \gamma_{h} + \gamma_j.
\end{eqnarray}

\section{Calculation of the soft functions}
In this appendix, we present the details of the calculation of the
two $\mathcal{O}(\alpha_s)$ soft functions $S^{(1)}_{bt}(k,\mu)$ and
$S^{(1)}_{tt}(k,\mu)$. We choose to do the calculation in the rest
frame of top quark, in which the  four-velocity of the top quark
is $v^{\mu}=(1,0,0,0)$. This choice of the frame makes the denominators
simple but leave the complexity in the delta functions.
Actually, we also perform the calculation in the frame where the delta
functions is simple but the singularities in the denominators are hard
to isolate~\cite{Kelley:2010qs}. And finally we obtain the same results,
which can be considered as a cross check for our calculations.

In the rest frame of top quark, we choose $n_b^{\mu}=(1,0,0,1)$. Then,
\begin{eqnarray}
q^{\mu}=q^+\frac{\bar{n}_b^{\mu}}{n_{b\bar{b}}}+q^-\frac{n_b^{\mu}}{n_{b\bar{b}}}+q_{\perp}^{\mu},\quad
n_1^{\mu}=n_1^+\frac{\bar{n}_b^{\mu}}{n_{b\bar{b}}}+n_1^{-}\frac{n_b^{\mu}}{n_{b\bar{b}}}+n_{1\perp}^{\mu},
\end{eqnarray}
and
\begin{eqnarray}
q\cdot n_1=\frac{q^+n_1^-+q^-n_1^+}{n_{b\bar{b}}}-|q_{\perp}||n_{1\perp}|\cos\theta,\quad
q\cdot v  =q \cdot \frac{(n_b+n_{\bar{b}})}{2}=\frac{(q^++q^-)}{2}.
\end{eqnarray}
Substituting these expressions into eq.~(\ref{eqs:soft}), we get
\begin{eqnarray}
S^{(1)}_{bt}(k,\mu)&=&\frac{g_s^2C_F\mu^{2\epsilon}}{(2\pi)^{d-1}} \int_0^{\infty}\hspace{-0.2cm}\mathrm{d}q^+\hspace{-0.1cm} \int_0^{\infty}\hspace{-0.2cm}\mathrm{d}q^-\hspace{-0.1cm}\int\mathrm{d}\Omega_{d-2}\left(\frac{2q^+q^-}{n_{b\bar{b}}} \right)^{-\epsilon}\nn\\
&&\delta(k-\frac{q^+n_1^-+q^-n_1^+}{n_{b\bar{b}}}+|q_{\perp}||n_{1\perp}|\cos\theta)\frac{n_b\cdot v}{q^+(q^++q^-)}.
\end{eqnarray}
Now redefine the integration variables $q^+$ and  $q^-$ and define $a=\frac{n_1^+}{n_1^-}$, then
\begin{eqnarray}
S^{(1)}_{bt}(k,\mu)&=&\frac{g_s^2C_F\mu^{2\epsilon}}{(2\pi)^{d-1}} \int_0^{\infty}\hspace{-0.2cm}\mathrm{d}q^+\hspace{-0.1cm} \int_0^{\infty}\hspace{-0.2cm}\mathrm{d}q^-\hspace{-0.1cm}\int\mathrm{d}\Omega_{d-2}\left(\frac{2n_{b\bar{b}}}{n_1^+n_1^-} \right)^{-\epsilon}\nn\\
&&\delta(k-q^+-q^-+2\sqrt{q^+q^-}\cos\theta)\frac{n_b\cdot v}{q^+(aq^++q^-)}.
\end{eqnarray}
Introducing two variables $x$ and $y$ such that $q^+=kyx$ and $q^-=ky(1-x)=ky\bar{x}$,
\begin{eqnarray}
S^{(1)}_{bt}(k,\mu)&=&\frac{g_s^2C_F\mu^{2\epsilon}}{(2\pi)^{d-1}}\left(\frac{2n_{b\bar{b}}}{n_1^+n_1^-} \right)^{-\epsilon} k^{-1-2\epsilon} \int\mathrm{d}\Omega_{d-2}\int_0^{1}\hspace{-0.2cm}\mathrm{d}x \nn\\
&&x^{-1-\epsilon}\frac{(1-2\sqrt{x\bar{x}}\cos\theta)^{2\epsilon}\bar{x}^{-\epsilon}}{ax+\bar{x}}.
\end{eqnarray}
The singularity in the integrand can be isolated by
\begin{eqnarray}
x^{-1-\epsilon}=-\frac{1}{\epsilon}\delta(x)+\left(\frac{1}{x}\right)_+-\epsilon\left(\frac{{\rm ln}x}{x}\right)_+ + \mathcal{O}(\epsilon^2).
\end{eqnarray}
Completing the above three parts of the integration separately and expanding
\begin{eqnarray}
\frac{1}{k^+}\left(\frac{\tilde{\mu}}{k^+}\right)^{2\epsilon}=-\frac{1}{2\epsilon}\delta(k^+)+
\left[\frac{1}{k^+}\right]_{\star}^{[k^+,\tilde{\mu}]}-2\epsilon\left[\frac{1}{k^+}{\rm ln}\frac{k^+}{\tilde{\mu}}\right]_{\star}^{[k^+,\tilde{\mu}]}+ \mathcal{O}(\epsilon^2),
\end{eqnarray}
we get
\begin{eqnarray}
S^{(1)}_{bt}(k,\mu)&=\frac{2C_F\alpha_s}{4\pi}\biggl\{ 4\biggl[ \frac{\ln \frac{k}{\tilde{\mu}}}{k} \biggr]_{\star}^{[k,\tilde{\mu}]}+\delta(k)c_{bt}^S \biggr\},
\end{eqnarray}
where
$c_{bt}^S=-{\rm ln}^2(1+\frac{1}{a})-2{\rm Li}_2(\frac{1}{1+a})+\frac{\pi^2}{12}$.

In a similar but simpler way, we can get $c_{tt}^S=-4{\rm ln}(1+\frac{1}{a})$.

\end{document}